\documentclass[%
reprint,
amsmath,amssymb,
aps,
]{revtex4-2}

\usepackage{graphicx}
\usepackage{dcolumn}
\usepackage{bm}

\begin{document}
		
	\preprint{APS/123-QED}
	
	\title{Comment on "Scalings of Rayleigh-Taylor Instability at Large Viscosity Contrasts\\ in Porous Media"}
	
	\author{E.B. Soboleva}
 \email{E-mail address: soboleva@ipmnet.ru}

\affiliation{%
	Ishlinsky Institute for Problems in Mechanics RAS,\\	
	Prospekt Vernadskogo 101-1, Moscow, 119526, Russia}
	
	\maketitle
	


%
%


In a recent Letter, Sabet et al. \cite{R01} presented new numerical results on a development of Rayleigh-Taylor (RT) instability in porous media. Two miscible fluids are assumed to be of different viscosity. The main finding is that a symmetry of gravity-driven convection is broken at large viscosity ratios $M=\mu_2^*/\mu_1^*>20$ and convective fingers extend preferably in the fluid of lower viscosity. If the upper fluid is much more viscous, as considered in the Letter, upward fingers extend at a much smaller rate than downward fingers. The authors try to explain the effect revealed by analyzing the vorticity field. The purpose of this Comment is to show that (1)  analyzing the vorticity field does not allow to explain this effect, (2) it can be explained by analyzing the velocity in the equation of motion. 

(1) Sabet et al. \cite{R01} decompose the vorticity field into density and viscosity counter-parts $\omega^{(\rho)}$ and $\omega^{(\mu)}$, respectively, and analyze their contributions. The authors treat the vorticity field as the driving force of convection that is not correct. They wrote, "The density component of vorticity $\omega^{(\rho)}$, which is the main driving force behind the growth of the fingers, is inversely proportional to the local value of viscosity. In other words, as the viscosity of the mixing layer increases, the driving force for buoyancy-driven convection is weakened..."  Actually, the fingers grow driven by the gravitational and buoyant forces. 

If the vorticity were able to cause the fingering asymmetry (according to the statement of Authors), it would influence the vertical translation of fluid. In this case, the vorticity has to change the vertical velocity of finger tips, which are the leading points of motion, and, particularly, to generate an additional negative vertical velocity of tips of upward fingers mitigating their motion. However, as clearly seen in Figure 1(b) in \cite{R01}, red and blue colors disappear at tips of both upward and downward fingers indicating that the value of vorticity is zero here. We see that the vorticity is absent at tips of both upward and downward fingers and therefore does not affect the vertical fluid velocity. One could assume that the additional vertical velocity of tip motion is generated by the vorticity near this tip. However, we do not find any explanation, why and in what way local rotations of fluid near the tip are transformed into an additional vertical translation of this tip. 

 \begin{figure}[!h]
	\begin{center}
		\includegraphics[width=1.0\columnwidth]{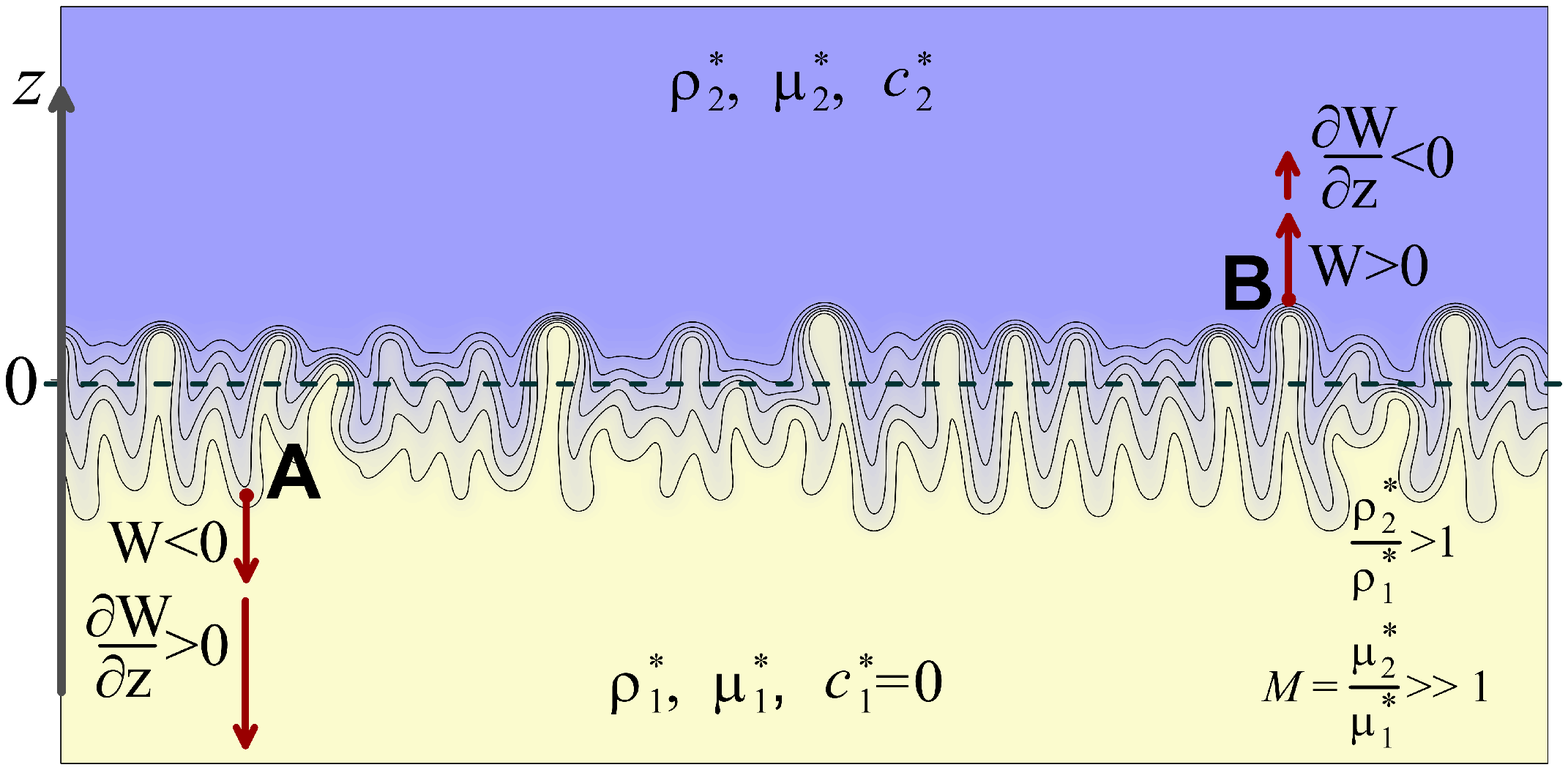}
		
		FIG. 1. Sketch of the problem.
	\end{center}
\end{figure}

(2) In \cite{R01}, fluid motions are described by the dimensionless Darcy equation [see Eq. (2)]. We transform this equation  replacing the Darcy velocity $\mathbf{v}$ with the fluid velocity $\mathbf{V}=\phi^{-1}\mathbf{v}$ [$\mathbf{V}=(U, W)$] and express the z-component of the velocity, $W$. One can denote $\partial p/\partial z+\rho(c)$ as $T$ and obtain the following:
\begin{equation}
	W=-\frac{1}{\phi \mu(c)} T, \  T= \frac{\partial p}{\partial z}+\rho(c) \ . \label{q004}
\end{equation}
The Cartesian coordinate system with the $z$ axis being vertical and oriented upward and all notations as in \cite{R01} are used. On the tip of downward finger, for example, at point $A$ in Fig. 1, fluid moves down at $W<0$ that leads to  $T>0$ here, according to Eq. (\ref{q004}); the values of $\phi$ and $\mu(c)$ are positive. On the tip of upward finger at point $B$, $W>0$, and $T<0$. 

We find the partial derivative of $W$ with respect to $z$:  
\begin{equation}
	\frac{\partial W}{\partial z}=-\frac{T}{\phi} \frac{\partial}{\partial z}\left( \frac{1}{\mu(c)}\right)-\frac{1}{\phi \mu(c)} \frac{\partial T}{\partial z} \ .     \label{q005}
\end{equation}
Further, the case of extremely large $M$ will be considered. In the mixed zone, the concentration $c$ changes from  $0$ to $1$  leading to a very large increase in the viscosity $\mu$ from $1$ to $M$. The viscosity changes per the unit of height are especially large on tips of convective fingers, therefore one can assume that the velocity variations are determined here mostly by the viscosity inhomogeneities. So, the second term on the right-hand side of Eq. (\ref{q005}) contributes less than the first term. It means that
\begin{equation}
 \arrowvert \frac{T}{\phi} \frac{\partial}{\partial z}\left( \frac{1}{\mu(c)}\right)\arrowvert >\arrowvert \frac{1}{\phi \mu(c)} \frac{\partial T}{\partial z}\arrowvert  \ .     \label{q006}
\end{equation}
Under the last assumption, the sign of the left-hand side of Eq. (\ref{q005}) coincides with the sign of the first term on the right-hand side, therefore, to analyze the sign of the derivative $\partial W/\partial z$, the second term can be neglected. We also transform the derivative of $1/\mu(c)$ and obtain
\begin{equation}
	\frac{\partial W}{\partial z}\approx T\frac{1}{\phi \mu^2(c)} \frac{\partial \mu(c)}{\partial c} \frac{\partial c}{\partial z}  \ .  \label{q007}
\end{equation}
 \vspace{1pt}
In Eq. (\ref{q007}), $\partial \mu(c)/\partial c>0$ because $\mu$ is a growing function of $c$, and $\partial c/\partial z>0$ because $c$ increases with height. Consequently, the signs of $\partial W/\partial z$ and $T$ coincide. At point $A$, $T>0$ leading to  $\partial W/\partial z>0$. Since $W<0$ here, $W$ becomes smaller or the absolute value $\mid W \mid $ becomes larger with decreasing $z$. In other words, if fluid moves down, its absolute velocity $\mid W \mid $ becomes larger. At point $B$, $T<0$ and $\partial W/\partial z<0$, that is $W$ becomes smaller if fluid moves up.

In the case of decreasing $\mu$ with  $c$, $\partial \mu(c)/\partial c<0$ and the signs of $\partial W/\partial z$ and $T$ are opposite. Further reasoning shows that, on the contrary, upward fingers have to accelerate and downward fingers have to decelerate. In summary, due to viscosity inhomogeneities, convective fingers extend preferably in the fluid of lower viscosity. This conclusion is consistent with one in \cite{R01} and confirmed by our own simulation \cite{R03}.
\vspace{4pt}

The study was supported by the Government program (contract AAAA-A20-120011690131-7).


\begin{thebibliography}{99}

\bibitem{R01}
N. Sabet, H. Hassanzadeh, A. De Wit, and J. Abedi. Scalings of Rayleigh-Taylor Instability at Large Viscosity Contrasts in Porous Media. Phys. Rev. Lett. {\bf126}, 094501 (2021). https://doi.org/10.1103/PhysRevLett.126.094501

\vspace{26pt}

\bibitem{R03}
E. B. Soboleva (unpublished).

\end{thebibliography}
\end{document}